\documentclass[aps,prl,amsmath,amssymb,reprint,longbibliography]{revtex4-1}

\usepackage{graphicx}
\usepackage{dcolumn}
\usepackage{bm}
\usepackage{upgreek}
\usepackage[colorlinks, linkcolor=blue, urlcolor=blue, anchorcolor=blue, citecolor=blue, dvipdfm]{hyperref}

\begin{document}


\title{All-optical Nanoscale Control of Photon Correlations: Dressed States Assisted Quantum Interference Effects}

\author{Dongxing Zhao$^1$}
\affiliation{$^1$School of Physical Science and Technology, Southwest University, Chongqing 400715, China}

\date{\today}

\begin{abstract}
  We propose an all-optical scheme to control the photon statistics using hybrid quantum plasmonic system. With the aid of dressed states assisted quantum interference effects, it is shown that the photon correlations of a signal field can be continuously modulated from bunching to antibunching under the control of a pump field. Apart from the exact multimode model, a single-mode model and an analytical treatment are also provided to help us identify the roles of multimode coupling and quantum interference between probability amplitudes. The proposed scheme, in contrast to the cavity quantum electrodynamics methods, works well even in the bad cavity limit. These findings suggest that this composite system provides a feasible nanophotonic platform for active modulation of photon statistics and for future quantum devices.
\end{abstract}


\maketitle

Due to excellent abilities to engineer the field-matter interaction, the hybrid quantum plasmonic system opens an alternative possibility to the realization of quantum-controlled devices \cite{Maier2013}. Among the rich variety of hybrid quantum plasmonic systems, the assembly of quantum emitter (QE) and metallic nanoparticle (MNP) plays a prominent role in the realization of coherent coupling between single quanta and emitters on the nanoscale. To explore its potential applications in the quantum science, various quantum optical properties of the hybrid QE-MNP system have been studied theoretically on the one hand, such as strong coupling \cite{Hohenester2008, Hughes2012, Rockstuhl2013, Vidal2014, Gu2017a, Xiao2017}, photon statistics \cite{Ridolfo2010, Rockstuhl2014a, Zhao2015, An2016, Vidal2017, Chen2017}, squeezing \cite{Agio2014, Cano2015, Carreno2017} and entanglement \cite{Angelakis2013, Rockstuhl2014b, Otten2015, Otten2016, Zubairy2016, Slowik2017, Gu2017b, Paspalakis2017}. On the other hand, remarkable experimental progresses about the Fano resonance \cite{Li2015, Zhang2017, Hou2017} and strong coupling \cite{Baumberg2016, Wang2017} of the QE-MNP system were reported. Though these rapid progresses promise the realization of more complex nanoscale quantum devices, the dependence of fabrication process or choice of dielectric environment to tune its optical response raises certain difficulties for practical applications. Thus it is highly desirable to develop the active way to control the quantum optical properties of QE-MNP system.

Characterized by equal-time second-order correlation function $g^{(2)}(0)$, the photon statistics exhibits the nonclassical features of light \cite{Scully1997}. A value of $g^{(2)}(0)<1$ ($> 1$) demonstrates the antibunching (bunching) statistics. The generation of nonclassical light fields, such as the antibunched light, is key to quantum networks \cite{Rempe2015} as well as quantum-optical spectroscopy \cite{Kira2011}. The traditional ways to tune the photon statistics generally rely on the cavity quantum electrodynamics (CQED) system which consists of QEs and optical microcavities. For instance, both abtibunched and bunched light can be obtained with the methods based on photon blockade and photon-induced tunneling effects \cite{Deutsch1997, Kimble2005}, cavity electromagnetically induced transparency \cite{Tan2002, Rempe2013}, or nonlinearity of coupled cavities \cite{Liew2010, Ciuti2011}. However, the dependence of strong coupling and micrometer size of CQED system sets up barriers for its ultracompact integrations and practical applications. Therefore, it is preferable to tune the photon statistics with a more feasible and compact scheme.

Here we investigate the all-optical modulation of photon correlations with the hybrid QE-MNP system. Different from the CQED system, the QE-MNP system can control the photon statistics without the need of strong coupling condition. To characterise the quantum nature of the studied system and perform in-depth study on this control scheme, we provide exact multimode quantum model which works beyond the dipole approximation. In addition, an approximate single-mode model and analytical expressions are also derived. It is found that enriched by the dressed states, the quantum interferences between the probability amplitudes play a vital role in the modulation of photon statistics.

\begin{figure}
  \includegraphics[width=8cm]{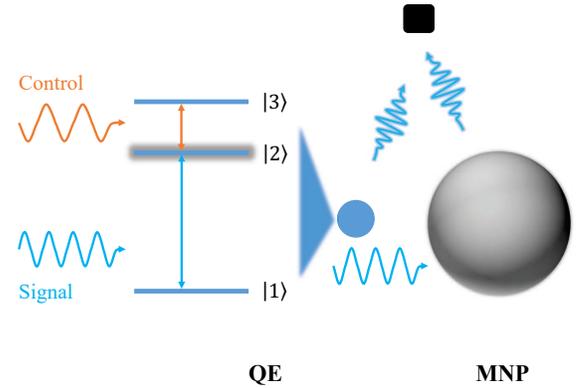}%
  \caption{\label{FIG-Schem} Schematic of the hybrid system composed of a QE and a spherical MNP. Both the dipole mode of MNP and QE transition $|1\rangle \leftrightarrow |2\rangle$ are driven by the signal field. In addition, the QE transition $|1\rangle \leftrightarrow |2\rangle$ couples to all the plasmon modes of MNP. A control field which drives the transition $|2\rangle \leftrightarrow |3\rangle$ is also introduced. }
\end{figure}

\emph{Exact multimode model.}---We consider a hybrid quantum plasmonic system comprised of a ladder-type three-level QE and spherical MNP with radius $r_{m}$ and dielectric constant $\varepsilon_{m}$, as shown in Fig. \ref{FIG-Schem}. The QE and MNP are separated by distance $R$ and embedded in a dielectric host (permittivity $\varepsilon_{b}$). MNP is able to support localized surface plasmon resonance, with the resonance frequency $\omega_{n}$ determined by $\textrm{Re}[ \varepsilon_{m} (\omega_{n})] = -\frac{n+1}{n} \varepsilon_{b}$ $(n=1, 2, ...)$ for spherical MNP whose radius is much smaller than the plasmon wavelength \cite{Bohren1983}. In contrast to the dipole mode ($n=1$), the higher modes ($n \geq 2$) cannot couple to an plane wave because of their vanishing dipole moment. The QE transition $|1\rangle \leftrightarrow |2\rangle$ (frequency $\omega_{21}$, decay rate $\gamma_{21}$, transition dipole moment $\mu$) is coupled by all the MNP modes as well as a weak signal field, with frequency $\omega_{s}$ and Rabi frequency $\Omega_{s\mu}$, while the transition $|2\rangle \leftrightarrow |3\rangle$ (frequency $\omega_{32}$, which is far-detuned from the plasmon frequencies, decay rate $\gamma_{32}$) is only coupled by a control field, with frequency $\omega_{c}$ and Rabi frequency $\Omega_{c}$. In addition, the dipole mode of MNP (dipole moment $\chi$) is also driven by the signal field, with Rabi frequency $\Omega_{s\chi}$. The Hamiltonian of the composite system in a rotating frame within the rotating-wave approximation reads ($\hbar=1$)
\begin{eqnarray} \label{EQU-EX}
  H &=& \Delta_{s}\sigma_{22}+(\Delta_{s}+\Delta_{c})\sigma_{33}+\sum_{n=1}^{N}\Delta_{n}a_{n}^{\dag}a_{n}\nonumber \\
    & & -\sum_{n=1}^{N}g_{n}(a_{n}\sigma_{21}+\textrm{H.c.}) \nonumber\\
    & & -(\Omega_{s\mu}\sigma_{12}+\Omega_{c}\sigma_{23}+\Omega_{s\chi}a_{1}+\textrm{H.c.}),
\end{eqnarray}
where $a_{n}$ ($a_{n}^{\dag}$) is the annihilation (creation) operator of of the MNP's $n$th mode, $\sigma_{ij}$ $(i,j=1, 2, 3)$ stands for a population operator for $i=j$ and a dipole transition operator for $i \neq j$, and H.c. represents Hermitian conjugate. The detunings are defined by $\Delta_{s} = \omega_{21} - \omega_{s}$, $\Delta_{c} = \omega_{32} - \omega_{c}$, and $\Delta_{n} = \omega_{n} - \omega_{s}$. Note that $N$ modes have been taken into consideration at most, which should be determined by the convergence of steady state results. The coupling strength between $n$th mode of MNP and QE transition $|1\rangle \leftrightarrow |2\rangle$ can be given by $g_{n} = \frac{\mu}{R^{n+2}} \sqrt{\frac{2n+1}{n} \frac{s_{n}\eta_{n}r_{m}^{2n+1}} {4\pi\hbar\varepsilon_{0}}}$, where $\eta_{n} = 1 / \frac{d}{d\omega} \textrm{Re} [\varepsilon_{m} (\omega)] |_{\omega=\omega_{n}}$ and $s_{n}=(n+1)^2$ $(n(n+1)/2)$ for a radial (tangential) QE \cite{Zhao2015}. As both the transition $|1\rangle \leftrightarrow |2\rangle$ and dipole mode of MNP are driven by the signal field, the ratio $\xi$ of $\Omega_{s\chi}$ and $\Omega_{s\mu}$ can be simply given by the ratio of $\chi$ and $\mu$, i.e. $\xi = \Omega_{s\chi}/\Omega_{s\mu} = \chi/\mu = \varepsilon_{b} \sqrt{ 12 \pi \hbar \varepsilon_{0} \eta_{1} r_{m}^{3}}/\mu$ \cite{Zhao2015}. The full dynamics of the system is governed by the master equation $\dot{\rho} = i[\rho,H] + \gamma_{21} \mathcal{L}[\sigma_{12}] \rho + \gamma_{32} \mathcal{L}[\sigma_{23}] \rho  + \sum_{n=1}^{N}\kappa_{n} \mathcal{L}[a_{n}] \rho$, where $\mathcal{L}[\hat{o}]\rho = \hat{o}\rho\hat{o}^{\dagger} - (\hat{o}^{\dagger} \hat{o} \rho + \rho\hat{o}^{\dagger} \hat{o})/2$ is the standard dissipator in Lindblad form; $\kappa_{n}= 2\eta_{n} \textrm{Im}[\varepsilon_{m} (\omega_{n})]$ is the decay rate of $n$th mode of the MNP. Under the conditions of weak pumping and small $N$, this equation can be numerically solved using the open-source software QuTiP \cite{Qutip2012, Qutip2013}.

\emph{Effective single-mode model.}---Though the multimode model gives the exact description of the studied system, the numerical calculations become unfeasible as the QE close to MNP, which may lead to exceedingly large dimensions of density matrix. In order to solve this problem and identify the role of higher-modes, we derive an equivalent effective Hamiltonian
\begin{eqnarray} \label{EQU-EF}
  H_{\textrm{eff}} &=& \Delta_{s,\textrm{eff}}\sigma_{22} + (\Delta_{s, \textrm{eff}}+\Delta_{c, \textrm{eff}})\sigma_{33} + \Delta_{1}a_{1}^{\dag}a_{1} \nonumber \\
    & & -g_{1}(a_{1}\sigma_{21}+\textrm{H.c.}) \nonumber\\
    & & -(\Omega_{s\mu}\sigma_{12}+\Omega_{c}\sigma_{23}+\Omega_{s\chi}a_{1}+\textrm{H.c.}),
\end{eqnarray}
and the corresponding effective master equation $\dot{\rho}_{\textrm{eff}} = i[\rho_{\textrm{eff}}, H_{\textrm{eff}}] + \gamma_{21,\textrm{eff}} \mathcal{L}[\sigma_{12}] \rho_{\textrm{eff}} + \gamma_{32} \mathcal{L}[\sigma_{23}] \rho_{\textrm{eff}}  + \kappa_{1} \mathcal{L}[a_{1}] \rho_{\textrm{eff}}$,
where the effective detunings $\Delta_{s,\textrm{eff}}$, $\Delta_{c,\textrm{eff}}$, and decay rates $\gamma_{21,\textrm{eff}}$ are modified as $\Delta_{s,\textrm{eff}} = \Delta_{s} - \sum_{n=2}^{N}\alpha_{n}(\omega_{n}-\omega_{x})$, $\Delta_{c,\textrm{eff}} = \Delta_{c} + \sum_{n=2}^{N}\alpha_{n}(\omega_{n}-\omega_{x})$, and $\gamma_{21,\textrm{eff}} = \gamma_{21} + \sum_{n=2}^{N}\alpha_{n}(\gamma_{n}-\gamma_{x})$, with $\alpha_{n}=\frac{g_{n}^{2}}{(\omega_{n}-\omega_{21})^2+(\kappa_{n}-\gamma_{21})^2/4}$. This effective model is derived with adiabatic elimination method, under the condition of $\alpha_{n} \ll 1$. For the realistic parameter, this approximation is valid as long as the QE is not in close proximity to MNP. This model not only greatly simplifies the calculation process, but also reveals that the roles of MNP higher-modes are to shift and broaden the level $|2\rangle$ of QE. To be more specific, the transition frequency between QE level $|1\rangle$ and $|2\rangle$ is modified to be $\omega_{21, \textrm{eff}}=\omega_{21} - \sum_{n=2}^{N}\alpha_{n}(\omega_{n}-\omega_{21})$ by higher modes of MNP.

\emph{Control of photon statistics}---To explore the nonclassical features of the system, we focus on the second-order correlation function $g^{(2)}(0)$ of the scattered signal field for the steady state, which can be calculated as $g^{(2)}(0)= \langle (\hat{P}^{\dag})^2 (\hat{P})^2 \rangle/\langle\hat{P}^{\dag}\hat{P}\rangle^2$, where $\hat{P}=\chi a_{1}+ \mu \sigma_{12}$ is the total polarization operator. With the polarization operator, the intensity can be calculated as $I=\langle \hat{P}^{\dag}\hat{P} \rangle$. We point out that one can also use input-output formalism to calculate $g^{(2)}(0)$ and $I$, which is essentially equivalent to the above expressions \cite{Waks2010}. In the following, we show that the quantum statistics can be controlled with the control field. As an example for proof-of-principle purposes, a silver MNP (radius $r_{m}=7$ nm) is consider hereafter, whose dielectric constants are given by the Drude model $\varepsilon_{m}(\omega)=\varepsilon_{\infty}-\frac{\omega_{p}^2}{\omega(\omega+i\gamma)}$, with the parameter being $\varepsilon_{\infty}=4.6$, $\omega_{p}=9.0$ eV, and $\gamma=0.1$ eV \cite{Vidal2014}. Furthermore, the parameters of QE are set to be $\mu=0.5$ enm, $\gamma_{21}=\gamma_{32}=0.05$ meV, and $\omega_{21}-\omega_{1}=-150$ meV. Placed in vacuum ($\varepsilon_{b}=1$), the QE and MNP are separated by a distance $R=12$ nm and driven by a weak signal field ($\Omega_{s}=0.005$ meV).

\begin{figure}
  \includegraphics[width=8cm]{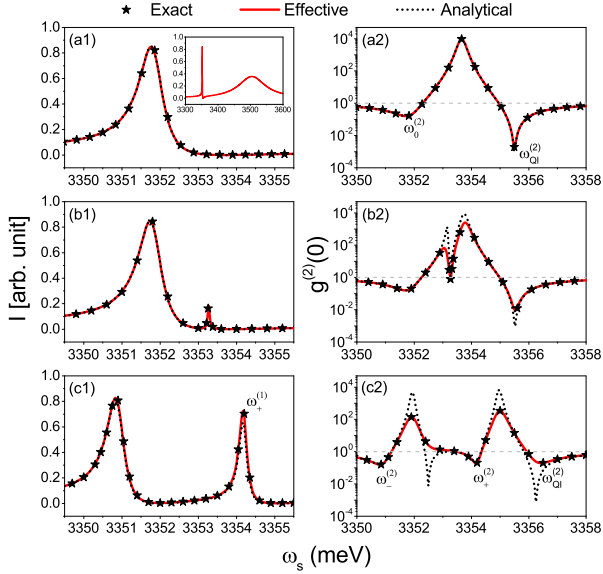}%
  \caption{\label{FIG-TSF}(Color online) Scattered intensity $I$ (a1-c1) and second-order correlation $g^{(2)}(0)$ (a2-c2) of the signal field versus the frequency $\omega_{s}$ for $\Omega_{c}=0$ (a), $\Omega_{c}=0.2$ meV (b), and $\Omega_{c}=1.5$ meV (c). Results of exact multimode model (solid stars), effective model (red solid curves), and analytical results (black dotted curves) are shown, respectively. The inset of (a1) displays the intensity spectra for a broader frequency when $\Omega_{c}=0$. The control field is set to be resonant with the transition $|2\rangle \leftrightarrow |3\rangle$ ($\Delta_{c} = 0$).}
\end{figure}

Figure \ref{FIG-TSF} shows the intensity $I$ and photon correlation $g^{(2)}(0)$ of the sinal field versus the frequency $\omega_{s}$. Results of effective model (red solid line) and multimode model (solid stars) are shown respectively, from which We can see excellent agreement between these two models. When the control field turns off [Fig. \ref{FIG-TSF}(a)], this hybrid system is just the same as a two-level QE-MNP system, which has been investigated in Ref. \cite{Zhao2015}. In this case, the scattered intensity spectrum [Fig. \ref{FIG-TSF}(a1)] displays a single-peaked structure, whose peak is located near the resonance frequency of QE transition $|1\rangle \leftrightarrow |2\rangle$. In fact, this structure is a Fano line shape, which can be clearly seen from the inset of Fig. \ref{FIG-TSF}(a1) displaying the intensity spectrum for a broader frequency $\omega_{s}$. This Fano structure originates from the coupling between the continuum excitations of MNP and the discrete excitations of QE \cite{Bryant2006}. Particularly, as shown in Fig. \ref{FIG-TSF}(a2), the photon statistics can be controlled from bunching to antibunching just by tuning the frequency $\omega_{s}$. Nevertheless, in order to realize the active control of photon correlation we still need to know the effects of control field.

To this end, the results when the control field turns on are shown in Fig. \ref{FIG-TSF}(b) ($\Omega_{c}=0.2$ meV) and Fig. \ref{FIG-TSF}(c) ($\Omega_{c}=1.5$ meV). For a relatively weak control field, we can see not only the different degree of the quantum correlation but also a new dip in $g^{(2)}(0)$ line shape [Fig. \ref{FIG-TSF}(b2)] . In addition, the intensity spectrum also shows a side peak centering at a frequency blue shifted from the main peak [Fig. \ref{FIG-TSF}(b1)]. Particularly, by further increasing the Rabi frequency of control field, the $g^{(2)}(0)$ shows more complex structures [Fig. \ref{FIG-TSF}(c2)]. There exist three frequency windows in which the scattered signal field shows antibunching statistics, which may provide versatile possibilities to control the quantum statistics. Furthermore, the two peaks in intensity spectrum for this case have nearly the same amplitudes [Fig. \ref{FIG-TSF}(c1)]. This line shape can be understood from the conception of dressed state \cite{Cohen1977}. In other words, dressed by the control field, the QE level $|2\rangle$ splits into a pair of well separated levels, whose energies are given by
\begin{equation}\label{EQU-DS}
 \omega_{\pm}^{(1)} = \omega_{21, \textrm{eff}} + \frac{1}{2}\left(\Delta_{c, \textrm{eff}} \pm \sqrt{\Delta_{c, \textrm{eff}}^{2} + 4 \Omega_{c}^{2}}\right).
\end{equation}
In the case discussed above, the $\omega_{-}^{(1)}$ is near the frequency of Fano peak. Therefore, both the $\omega_{-}^{(1)}$ level and the Fano interference contribute to the low-energy peak, and thus this peak frequency is shifted compared with $\omega_{-}^{(1)}$ and the main peak for $\Omega_{c}=0$. For the high-energy peak, its peak frequency is just located at $\omega_{+}^{(1)}$.

\emph{Mechanism of quantum interference.}---In order to gain deeper physical insights into the above results, we present an analytical description for the studied system. Under weak pumping conditions, the state of the composite system can be approximated as a pure state \cite{Carmichael1991, Ciuti2011, Vidal2017}, whose evolution is governed by Schr\"{o}dinger equation. After truncating the Hilbert space to two excitations, the state of the hybrid system in the rotating frame is given by $|\psi\rangle = |0,1\rangle + C_{0,2}|0,2\rangle + C_{0,3}|0,3\rangle + C_{1,1}|1,1\rangle + C_{1,2}|1,2\rangle + C_{1,3}|1,3\rangle + C_{2,1}|2,1\rangle$, where $|m,n\rangle$ denotes the plasmon Fock state $|m\rangle$ and QE level $|n\rangle$. The system dissipations can be included by considering a non-Hermitian Hamiltonian
\begin{equation}
 H_{\textrm{eff}}^{\textrm{nH}} = H_{\textrm{eff}} - i\frac{\gamma_{21,\textrm{eff}}}{2}\sigma_{22} -i\frac{\gamma_{32}}{2}\sigma_{33}
 -i\frac{\kappa_{1}}{2}a_{1}^{\dag}a_{1}.
\end{equation}
Then the scattered intensity $I$ is derived as
\begin{eqnarray}\label{EQU-ANAI}
  I &=& \mu^2 (|C_{0,2}|^{2} + \xi^{2} |C_{1,1}|^{2} + 2\xi \textrm{Re}[C_{0,2}^{*}C_{1,1}]) \nonumber \\
  &=& \mu^2\Omega_{s}^{2}\left|\frac{\widetilde{\Delta}_{1} + \xi^{2} \left(\widetilde{\Delta}_{s,\textrm{eff}} - \frac{\Omega_{c}^{2}}{\widetilde{\Delta}_{sc,\textrm{eff}}}\right) + 2 \xi g_{1}}{\widetilde{\Delta}_{1}\left(\widetilde{\Delta}_{s,\textrm{eff}} - \frac{\Omega_{c}^{2}}{\widetilde{\Delta}_{sc,\textrm{eff}}}\right)-g_{1}^{2}}\right|^2,
\end{eqnarray}
where $\widetilde{\Delta}_{1}=\Delta_{1}-i\kappa_{1}/2$, $\widetilde{\Delta}_{s,\textrm{eff}}=\Delta_{s,\textrm{eff}}-i\gamma_{21,\textrm{eff}}/2$, and $\widetilde{\Delta}_{sc,\textrm{eff}}=\Delta_{s,\textrm{eff}}+\Delta_{c,\textrm{eff}} - i\gamma_{32}/2$. Moreover, the second-order correlation function can be determined by
\begin{equation}\label{EQU-ANAg21}
 g^{(2)}(0) = \frac{2 \xi^{2} \left(2|C_{1,2}|^{2} + \xi^{2} |C_{2,1}|^{2} + 2\sqrt{2}\xi \textrm{Re}[C_{1,2}^{*}C_{2,1}]\right)}{\left[|C_{0,2}|^{2} + \xi^{2} |C_{1,1}|^{2} + 2\xi \textrm{Re}[C_{0,2}^{*}C_{1,1}]\right]^{2}},
\end{equation}
whose full expression is found to be
\begin{widetext}
\begin{equation}\label{EQU-ANAg22}
 g^{(2)}(0) = \left| 1-\frac{(\widetilde{\Delta}_{1} + \xi g_{1})^2\left[\widetilde{\Delta}_{1}  \left(\widetilde{\Delta}_{s,\textrm{eff}} - \frac{\Omega_{c}^{2}} {\widetilde{\Delta}_{sc,\textrm{eff}}} \right) - g_{1}^{2} + (\widetilde{\Delta}_{1} + \xi g_{1})^2\left(1 + \frac{\Omega_{c}^{2}}{\widetilde{\Delta}_{sc,\textrm{eff}}(\widetilde{\Delta}_{1} +\widetilde{\Delta}_{sc,\textrm{eff}})}\right)\right]}{\left[\widetilde{\Delta}_{1} \left(\widetilde{\Delta}_{1} + \widetilde{\Delta}_{s,\textrm{eff}} - \frac{\Omega_{c}^{2}}{\widetilde{\Delta}_{1} + \widetilde{\Delta}_{sc,\textrm{eff}}}\right) - g_{1}^{2}\right]\left[\widetilde{\Delta}_{1} + 2 \xi g_{1} + \xi^{2} \left(\widetilde{\Delta}_{s,\textrm{eff}} - \frac{\Omega_{c}^{2}}{\widetilde{\Delta}_{sc,\textrm{eff}}}\right)\right]^2}\right|^2.
\end{equation}
\end{widetext}
The validity of these analytical results is demonstrated in Fig. \ref{FIG-TSF} (dotted lines).

\begin{figure}
  \includegraphics[width=8cm]{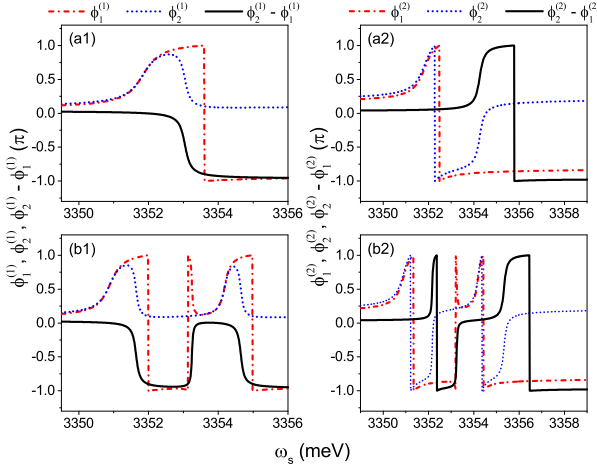}%
  \caption{\label{FIG-PHA}(Color online) Phase parameters $\phi^{(1)}_{1}$, $\phi^{(1)}_{2}$, phase difference $\phi^{(1)}_{2} - \phi^{(1)}_{1}$ (a1, b1) and phase parameters $\phi^{(2)}_{1}$, $\phi^{(2)}_{2}$ and their difference $\phi^{(2)}_{2} - \phi^{(2)}_{1}$ as a function of $\omega_{s}$ for the case of $\Omega_{c}=0$ (a1, a2) and $\Omega_{c}=1.5$ meV (b1, b2).}
\end{figure}

The analytical results illustrate the roles of the interference of probability amplitudes. Clear quantum interference signatures are shown in Eqs. (\ref{EQU-ANAI}) and (\ref{EQU-ANAg21}). For the intensity spectrum, the interference of $C_{0,2}$ and $C_{1,1}$, which are the probability amplitudes for one-quantum states $|0,2\rangle$ and $|1,1\rangle$, determines the line shape of intensity spectrum. By contrast, the interference between probability amplitudes of two-quantum states $|1,2\rangle$ and $|2,1\rangle$ determines the second-order correlation. In order to study the underlying quantum interference effects in depth, we introduce four phase parameters ($\phi^{(1)}_{1}, \phi^{(1)}_{2}, \phi^{(2)}_{1}, \phi^{(2)}_{2}$), which are the arguments of four complex probability amplitudes ($C_{0,2}, C_{1,1}, C_{1,2}, C_{2,1}$). It is obvious that the phase difference $\phi^{(1)}_{2} - \phi^{(1)}_{1}$ determines the interference properties of one-quantum amplitudes, while the phase difference $\phi^{(2)}_{2} - \phi^{(2)}_{1}$ gives the quantum interference effects on second-order correlation.

Figure \ref{FIG-PHA} displays the phase parameters versus the frequency of signal field for the case of $\Omega_{c}=0$ (a) and $\Omega_{c}=1.5$ meV (b). When the control field is absent, the interference properties are relatively simple. The first-order phase $\phi^{(1)}_{1}$ and $\phi^{(1)}_{2}$ interference constructively (destructively) below (above) the frequency $\omega_{21,\textrm{eff}}$. It is the transition from constructive to destructive interference that contributes to the one-peaked Fano structure in Fig. \ref{FIG-TSF}(a1). However, after exerting a strong control field, the QE level $|2\rangle$ splits into two dressed states, which leads to twice transitions from constructive to destructive interference [Fig. \ref{FIG-PHA}(b1)]. Hence, the scattered spectrum for $\Omega_{c}=1.5$ meV shows two peaks.

Next we focus on the quantum interference between two-quantum probability amplitudes. Without the control field, the amplitudes $C_{1,2}$ and $C_{2,1}$ interference destructively when $\omega_{s}>3354$ meV. For an appropriate frequency, the third term in the numerator of Eq. (\ref{EQU-ANAg21}) cancels the first two terms, which leads to the photon antibunching in Fig. \ref{FIG-TSF}(a2) around frequency $\omega_{\textrm{QI}}^{(2)}$. This particular interference effect is induced by the openness of plasmonic cavity. Although the amplitudes $C_{1,2}$ and $C_{2,1}$ interference constructively when $\omega_{s}<3354$  meV, we can still observe photon antibunching effect around frequency $\omega_{0}^{(2)}$. We point out that this antibunching originates from the photon blockade effect \cite{Kimble2005}. By diagonalizing the Eq. (\ref{EQU-EF}) with the absence of driven fields, the $\omega_{0}^{(2)}$ is derived as
\begin{equation}
 \omega_{0}^{(2)} = \omega_{1} + \frac{1}{2}\left[\omega_{21,\textrm{eff}} - \omega_{1} - \sqrt{(\omega_{21,\textrm{eff}} - \omega_{1})^{2} + 4 g_{1}^{2} }\right].
\end{equation}
For the case with strong control field, three antibunching regions can be found. According to the phase difference between $\phi^{(2)}_{1}$ and $\phi^{(2)}_{2}$, the antibunching around $\omega_{\textrm{QI}}^{(2)}$ in Fig. \ref{FIG-TSF}(c2) is also induced by the destructive quantum interference. In contrast, there are two antibunching regions (denoted as $\omega_{\pm}^{(2)}$) induced by photon blockade effect for this case. Using the dressed states energy in Eq. (\ref{EQU-DS}), the $\omega_{\pm}^{(2)}$ read
\begin{equation}
 \omega_{\pm}^{(2)} = \omega_{1} + \frac{1}{2}\left[\omega_{\pm}^{(1)} - \omega_{1} - \sqrt{(\omega_{\pm}^{(1)} - \omega_{1})^{2} + 4 g_{1}^{2} }\right].
\end{equation}

\begin{figure}
  \includegraphics[width=8cm]{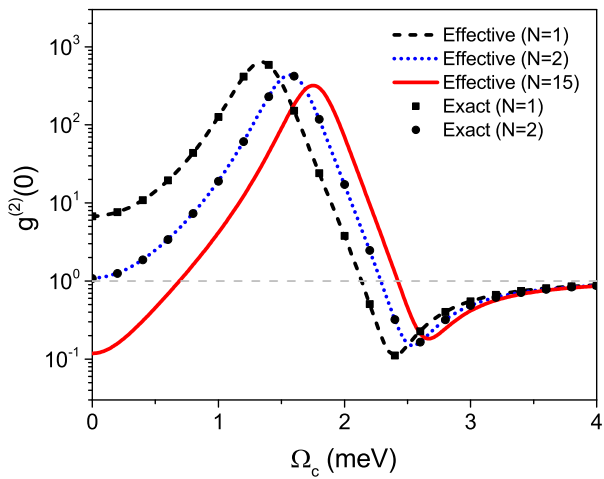}%
  \caption{\label{FIG-TCI}(Color online) $g^{(2)}(0)$ correlation with respect to the Rabi frequency $\Omega_{c}$ calculated from effective model (red solid). The results of $N=1$ and $N=2$ calculated from effective model (dashed and dotted lines) and exact model (scatterers) are also shown. The detuning $\Delta_{s}$ is fixed to $\Delta_{s}=-2$ meV.}
\end{figure}

So far we have shown the possibility of the photon correlation modulation by light field and the underlying quantum interference mechanism. In real applications, it is more desirable to realize the continuous modulation between bunching and antibunching. In the following, we show the all-optical modulation of photon correlation with the Rabi frequency $\Omega_{c}$ in Fig. \ref{FIG-TCI}. When the Rabi frequency $\Omega_{c}$ is increased, the photon correlation of signal field makes a continuous transition from antibunching to bunching, and then to antibunching. The photon correlation tends to disappear, which means a field with random statistics, for strong control field ($\Omega_{c}>4$ meV). Note that the coupling between QE and MNP is not in strong coupling regime for the case studied here. However, the peculiar quantum interference effects enable us to control the photon correlations even in the bad cavity limit. Therefore, this result represents a nanoscaled scheme for all-optical control of photon statistics without the need of strong coupling condition. Compared with the methods in CQED, the scheme presented here is more compact and feasible. By the way, we point out that the impact of high-order modes of MNP is non-ignorable when the distance between QE and MNP is small. Neglecting the contribution of high-order modes may lead to a contrary prediction about the photon statistics [see the difference between solid and dashed curves in Fig. \ref{FIG-TCI}].

\emph{Conclusions.}---To summarize, we have theoretically demonstrated the all-optical control of photon statistics with a hybrid QE-MNP system. We present three different ways for the studied system to uncover the nonclassical correlation and underlying mechanisms. The analytical treatment reveals that the quantum interferences between two-quantum states' amplitudes are essential to the photon correlation modulation. Enriched by the dressed states, this quantum interference effect enables us to continuously regulate the photon correlation between bunching and antibunching. Compared with the CQED methods, this scheme is more compact and experimentally feasible as it works in the bad cavity limit. These results open an alternative possibility for the control of photon correlation, which may find its applications in active quantum-optical devices.


\begin{acknowledgments}
This work was supported by the Fundamental Research Funds for the Central Universities of China under Grant No. SWU116056.
\end{acknowledgments}

\bibliography{AOCP-ref}

\begin{thebibliography}{45}%
\makeatletter
\providecommand \@ifxundefined [1]{%
 \@ifx{#1\undefined}
}%
\providecommand \@ifnum [1]{%
 \ifnum #1\expandafter \@firstoftwo
 \else \expandafter \@secondoftwo
 \fi
}%
\providecommand \@ifx [1]{%
 \ifx #1\expandafter \@firstoftwo
 \else \expandafter \@secondoftwo
 \fi
}%
\providecommand \natexlab [1]{#1}%
\providecommand \enquote  [1]{``#1''}%
\providecommand \bibnamefont  [1]{#1}%
\providecommand \bibfnamefont [1]{#1}%
\providecommand \citenamefont [1]{#1}%
\providecommand \href@noop [0]{\@secondoftwo}%
\providecommand \href [0]{\begingroup \@sanitize@url \@href}%
\providecommand \@href[1]{\@@startlink{#1}\@@href}%
\providecommand \@@href[1]{\endgroup#1\@@endlink}%
\providecommand \@sanitize@url [0]{\catcode `\\12\catcode `\$12\catcode
  `\&12\catcode `\#12\catcode `\^12\catcode `\_12\catcode `\%12\relax}%
\providecommand \@@startlink[1]{}%
\providecommand \@@endlink[0]{}%
\providecommand \url  [0]{\begingroup\@sanitize@url \@url }%
\providecommand \@url [1]{\endgroup\@href {#1}{\urlprefix }}%
\providecommand \urlprefix  [0]{URL }%
\providecommand \Eprint [0]{\href }%
\providecommand \doibase [0]{http://dx.doi.org/}%
\providecommand \selectlanguage [0]{\@gobble}%
\providecommand \bibinfo  [0]{\@secondoftwo}%
\providecommand \bibfield  [0]{\@secondoftwo}%
\providecommand \translation [1]{[#1]}%
\providecommand \BibitemOpen [0]{}%
\providecommand \bibitemStop [0]{}%
\providecommand \bibitemNoStop [0]{.\EOS\space}%
\providecommand \EOS [0]{\spacefactor3000\relax}%
\providecommand \BibitemShut  [1]{\csname bibitem#1\endcsname}%
\let\auto@bib@innerbib\@empty
\bibitem [{\citenamefont {Tame}\ \emph {et~al.}(2013)\citenamefont {Tame},
  \citenamefont {McEnery}, \citenamefont {Ozdemir}, \citenamefont {Lee},
  \citenamefont {Maier},\ and\ \citenamefont {Kim}}]{Maier2013}%
  \BibitemOpen
  \bibfield  {author} {\bibinfo {author} {\bibfnamefont {M.~S.}\ \bibnamefont
  {Tame}}, \bibinfo {author} {\bibfnamefont {K.~R.}\ \bibnamefont {McEnery}},
  \bibinfo {author} {\bibfnamefont {S.~K.}\ \bibnamefont {Ozdemir}}, \bibinfo
  {author} {\bibfnamefont {J.}~\bibnamefont {Lee}}, \bibinfo {author}
  {\bibfnamefont {S.~A.}\ \bibnamefont {Maier}}, \ and\ \bibinfo {author}
  {\bibfnamefont {M.~S.}\ \bibnamefont {Kim}},\ }\bibfield  {title} {\enquote
  {\bibinfo {title} {Quantum plasmonics},}\ }\href
  {https://doi.org/10.1038/nphys2615} {\bibfield  {journal} {\bibinfo
  {journal} {Nat.\ Phys.}\ }\textbf {\bibinfo {volume} {9}},\ \bibinfo {pages}
  {329} (\bibinfo {year} {2013})}\BibitemShut {NoStop}%
\bibitem [{\citenamefont {Tr\"ugler}\ and\ \citenamefont
  {Hohenester}(2008)}]{Hohenester2008}%
  \BibitemOpen
  \bibfield  {author} {\bibinfo {author} {\bibfnamefont {A.}~\bibnamefont
  {Tr\"ugler}}\ and\ \bibinfo {author} {\bibfnamefont {U.}~\bibnamefont
  {Hohenester}},\ }\bibfield  {title} {\enquote {\bibinfo {title} {Strong
  coupling between a metallic nanoparticle and a single molecule},}\ }\href
  {\doibase 10.1103/PhysRevB.77.115403} {\bibfield  {journal} {\bibinfo
  {journal} {Phys. Rev. B}\ }\textbf {\bibinfo {volume} {77}},\ \bibinfo
  {pages} {115403} (\bibinfo {year} {2008})}\BibitemShut {NoStop}%
\bibitem [{\citenamefont {Van~Vlack}\ \emph {et~al.}(2012)\citenamefont
  {Van~Vlack}, \citenamefont {Kristensen},\ and\ \citenamefont
  {Hughes}}]{Hughes2012}%
  \BibitemOpen
  \bibfield  {author} {\bibinfo {author} {\bibfnamefont {C.}~\bibnamefont
  {Van~Vlack}}, \bibinfo {author} {\bibfnamefont {P.~T.}\ \bibnamefont
  {Kristensen}}, \ and\ \bibinfo {author} {\bibfnamefont {S.}~\bibnamefont
  {Hughes}},\ }\bibfield  {title} {\enquote {\bibinfo {title} {Spontaneous
  emission spectra and quantum light-matter interactions from a strongly
  coupled quantum dot metal-nanoparticle system},}\ }\href {\doibase
  10.1103/PhysRevB.85.075303} {\bibfield  {journal} {\bibinfo  {journal} {Phys.
  Rev. B}\ }\textbf {\bibinfo {volume} {85}},\ \bibinfo {pages} {075303}
  (\bibinfo {year} {2012})}\BibitemShut {NoStop}%
\bibitem [{\citenamefont {S\l{}owik}\ \emph {et~al.}(2013)\citenamefont
  {S\l{}owik}, \citenamefont {Filter}, \citenamefont {Straubel}, \citenamefont
  {Lederer},\ and\ \citenamefont {Rockstuhl}}]{Rockstuhl2013}%
  \BibitemOpen
  \bibfield  {author} {\bibinfo {author} {\bibfnamefont {K.}~\bibnamefont
  {S\l{}owik}}, \bibinfo {author} {\bibfnamefont {R.}~\bibnamefont {Filter}},
  \bibinfo {author} {\bibfnamefont {J.}~\bibnamefont {Straubel}}, \bibinfo
  {author} {\bibfnamefont {F.}~\bibnamefont {Lederer}}, \ and\ \bibinfo
  {author} {\bibfnamefont {C.}~\bibnamefont {Rockstuhl}},\ }\bibfield  {title}
  {\enquote {\bibinfo {title} {Strong coupling of optical nanoantennas and
  atomic systems},}\ }\href {\doibase 10.1103/PhysRevB.88.195414} {\bibfield
  {journal} {\bibinfo  {journal} {Phys. Rev. B}\ }\textbf {\bibinfo {volume}
  {88}},\ \bibinfo {pages} {195414} (\bibinfo {year} {2013})}\BibitemShut
  {NoStop}%
\bibitem [{\citenamefont {Delga}\ \emph {et~al.}(2014)\citenamefont {Delga},
  \citenamefont {Feist}, \citenamefont {Bravo-Abad},\ and\ \citenamefont
  {Garcia-Vidal}}]{Vidal2014}%
  \BibitemOpen
  \bibfield  {author} {\bibinfo {author} {\bibfnamefont {A.}~\bibnamefont
  {Delga}}, \bibinfo {author} {\bibfnamefont {J.}~\bibnamefont {Feist}},
  \bibinfo {author} {\bibfnamefont {J.}~\bibnamefont {Bravo-Abad}}, \ and\
  \bibinfo {author} {\bibfnamefont {F.~J.}\ \bibnamefont {Garcia-Vidal}},\
  }\bibfield  {title} {\enquote {\bibinfo {title} {Quantum emitters near a
  metal nanoparticle: Strong coupling and quenching},}\ }\href {\doibase
  10.1103/PhysRevLett.112.253601} {\bibfield  {journal} {\bibinfo  {journal}
  {Phys. Rev. Lett.}\ }\textbf {\bibinfo {volume} {112}},\ \bibinfo {pages}
  {253601} (\bibinfo {year} {2014})}\BibitemShut {NoStop}%
\bibitem [{\citenamefont {Ren}\ \emph {et~al.}(2017)\citenamefont {Ren},
  \citenamefont {Gu}, \citenamefont {Zhao}, \citenamefont {Zhang},
  \citenamefont {Zhang},\ and\ \citenamefont {Gong}}]{Gu2017a}%
  \BibitemOpen
  \bibfield  {author} {\bibinfo {author} {\bibfnamefont {J.}~\bibnamefont
  {Ren}}, \bibinfo {author} {\bibfnamefont {Y.}~\bibnamefont {Gu}}, \bibinfo
  {author} {\bibfnamefont {D.}~\bibnamefont {Zhao}}, \bibinfo {author}
  {\bibfnamefont {F.}~\bibnamefont {Zhang}}, \bibinfo {author} {\bibfnamefont
  {T.}~\bibnamefont {Zhang}}, \ and\ \bibinfo {author} {\bibfnamefont
  {Q.}~\bibnamefont {Gong}},\ }\bibfield  {title} {\enquote {\bibinfo {title}
  {Evanescent-vacuum-enhanced photon-exciton coupling and fluorescence
  collection},}\ }\href {\doibase 10.1103/PhysRevLett.118.073604} {\bibfield
  {journal} {\bibinfo  {journal} {Phys. Rev. Lett.}\ }\textbf {\bibinfo
  {volume} {118}},\ \bibinfo {pages} {073604} (\bibinfo {year}
  {2017})}\BibitemShut {NoStop}%
\bibitem [{\citenamefont {Peng}\ \emph {et~al.}(2017)\citenamefont {Peng},
  \citenamefont {Liu}, \citenamefont {Xu}, \citenamefont {Cao}, \citenamefont
  {Lu}, \citenamefont {Gong},\ and\ \citenamefont {Xiao}}]{Xiao2017}%
  \BibitemOpen
  \bibfield  {author} {\bibinfo {author} {\bibfnamefont {P.}~\bibnamefont
  {Peng}}, \bibinfo {author} {\bibfnamefont {Y.-C.}\ \bibnamefont {Liu}},
  \bibinfo {author} {\bibfnamefont {D.}~\bibnamefont {Xu}}, \bibinfo {author}
  {\bibfnamefont {Q.-T.}\ \bibnamefont {Cao}}, \bibinfo {author} {\bibfnamefont
  {G.}~\bibnamefont {Lu}}, \bibinfo {author} {\bibfnamefont {Q.}~\bibnamefont
  {Gong}}, \ and\ \bibinfo {author} {\bibfnamefont {Y.-F.}\ \bibnamefont
  {Xiao}},\ }\bibfield  {title} {\enquote {\bibinfo {title} {Enhancing coherent
  light-matter interactions through microcavity-engineered plasmonic
  resonances},}\ }\href {https://doi.org/10.1103/PhysRevLett.119.233901}
  {\bibfield  {journal} {\bibinfo  {journal} {Phys. Rev. Lett.}\ }\textbf
  {\bibinfo {volume} {119}},\ \bibinfo {pages} {233901} (\bibinfo {year}
  {2017})}\BibitemShut {NoStop}%
\bibitem [{\citenamefont {Ridolfo}\ \emph {et~al.}(2010)\citenamefont
  {Ridolfo}, \citenamefont {DiStefano}, \citenamefont {Fina}, \citenamefont
  {Saija},\ and\ \citenamefont {Savasta}}]{Ridolfo2010}%
  \BibitemOpen
  \bibfield  {author} {\bibinfo {author} {\bibfnamefont {A.}~\bibnamefont
  {Ridolfo}}, \bibinfo {author} {\bibfnamefont {O.}~\bibnamefont {DiStefano}},
  \bibinfo {author} {\bibfnamefont {N.}~\bibnamefont {Fina}}, \bibinfo {author}
  {\bibfnamefont {R.}~\bibnamefont {Saija}}, \ and\ \bibinfo {author}
  {\bibfnamefont {S.}~\bibnamefont {Savasta}},\ }\bibfield  {title} {\enquote
  {\bibinfo {title} {Quantum plasmonics with quantum dot-metal nanoparticle
  molecules: Influence of the fano effect on photon statistics},}\ }\href
  {https://doi.org/10.1103/PhysRevLett.105.263601} {\bibfield  {journal}
  {\bibinfo  {journal} {Phys.\ Rev.\ Lett.}\ }\textbf {\bibinfo {volume}
  {105}},\ \bibinfo {pages} {263601} (\bibinfo {year} {2010})}\BibitemShut
  {NoStop}%
\bibitem [{\citenamefont {Filter}\ \emph {et~al.}(2014)\citenamefont {Filter},
  \citenamefont {S\l{}owik}, \citenamefont {Straubel}, \citenamefont
  {Lederer},\ and\ \citenamefont {Rockstuhl}}]{Rockstuhl2014a}%
  \BibitemOpen
  \bibfield  {author} {\bibinfo {author} {\bibfnamefont {R.}~\bibnamefont
  {Filter}}, \bibinfo {author} {\bibfnamefont {K.}~\bibnamefont {S\l{}owik}},
  \bibinfo {author} {\bibfnamefont {J.}~\bibnamefont {Straubel}}, \bibinfo
  {author} {\bibfnamefont {F.}~\bibnamefont {Lederer}}, \ and\ \bibinfo
  {author} {\bibfnamefont {C.}~\bibnamefont {Rockstuhl}},\ }\bibfield  {title}
  {\enquote {\bibinfo {title} {Nanoantennas for ultrabright single photon
  sources},}\ }\href {https://dx.doi.org/10.1364/OL.39.001246} {\bibfield
  {journal} {\bibinfo  {journal} {Opt. Lett.}\ }\textbf {\bibinfo {volume}
  {39}},\ \bibinfo {pages} {1246} (\bibinfo {year} {2014})}\BibitemShut
  {NoStop}%
\bibitem [{\citenamefont {Zhao}\ \emph {et~al.}(2015)\citenamefont {Zhao},
  \citenamefont {Gu}, \citenamefont {Chen}, \citenamefont {Ren}, \citenamefont
  {Zhang},\ and\ \citenamefont {Gong}}]{Zhao2015}%
  \BibitemOpen
  \bibfield  {author} {\bibinfo {author} {\bibfnamefont {D.}~\bibnamefont
  {Zhao}}, \bibinfo {author} {\bibfnamefont {Y.}~\bibnamefont {Gu}}, \bibinfo
  {author} {\bibfnamefont {H.}~\bibnamefont {Chen}}, \bibinfo {author}
  {\bibfnamefont {J.}~\bibnamefont {Ren}}, \bibinfo {author} {\bibfnamefont
  {T.}~\bibnamefont {Zhang}}, \ and\ \bibinfo {author} {\bibfnamefont
  {Q.}~\bibnamefont {Gong}},\ }\bibfield  {title} {\enquote {\bibinfo {title}
  {Quantum statistics control with a plasmonic nanocavity: Multimode-enhanced
  interferences},}\ }\href {https://doi.org/10.1103/PhysRevA.92.033836}
  {\bibfield  {journal} {\bibinfo  {journal} {Phys. Rev. A}\ }\textbf {\bibinfo
  {volume} {92}},\ \bibinfo {pages} {033836} (\bibinfo {year}
  {2015})}\BibitemShut {NoStop}%
\bibitem [{\citenamefont {Yang}\ and\ \citenamefont {An}(2016)}]{An2016}%
  \BibitemOpen
  \bibfield  {author} {\bibinfo {author} {\bibfnamefont {C.-J.}\ \bibnamefont
  {Yang}}\ and\ \bibinfo {author} {\bibfnamefont {J.-H.}\ \bibnamefont {An}},\
  }\bibfield  {title} {\enquote {\bibinfo {title} {Resonance fluorescence
  beyond the dipole approximation of a quantum dot in a plasmonic
  nanostructure},}\ }\href {\doibase 10.1103/PhysRevA.93.053803} {\bibfield
  {journal} {\bibinfo  {journal} {Phys. Rev. A}\ }\textbf {\bibinfo {volume}
  {93}},\ \bibinfo {pages} {053803} (\bibinfo {year} {2016})}\BibitemShut
  {NoStop}%
\bibitem [{\citenamefont {S\'{a}ez-Bl\'{a}zquez}\ \emph
  {et~al.}(2017)\citenamefont {S\'{a}ez-Bl\'{a}zquez}, \citenamefont {Feist},
  \citenamefont {Fern\'{a}ndez-Dom\'{\i}nguez},\ and\ \citenamefont
  {Garc\'{\i}a-Vidal}}]{Vidal2017}%
  \BibitemOpen
  \bibfield  {author} {\bibinfo {author} {\bibfnamefont {R.}~\bibnamefont
  {S\'{a}ez-Bl\'{a}zquez}}, \bibinfo {author} {\bibfnamefont {J.}~\bibnamefont
  {Feist}}, \bibinfo {author} {\bibfnamefont {A.~I.}\ \bibnamefont
  {Fern\'{a}ndez-Dom\'{\i}nguez}}, \ and\ \bibinfo {author} {\bibfnamefont
  {F.~J.}\ \bibnamefont {Garc\'{\i}a-Vidal}},\ }\bibfield  {title} {\enquote
  {\bibinfo {title} {Enhancing photon correlations through plasmonic strong
  coupling},}\ }\href {https://doi.org/10.1364/OPTICA.4.001363} {\bibfield
  {journal} {\bibinfo  {journal} {Optica}\ }\textbf {\bibinfo {volume} {4}},\
  \bibinfo {pages} {1363} (\bibinfo {year} {2017})}\BibitemShut {NoStop}%
\bibitem [{\citenamefont {Zhang}\ \emph
  {et~al.}(2017{\natexlab{a}})\citenamefont {Zhang}, \citenamefont {Protsenko},
  \citenamefont {Sandoghdar},\ and\ \citenamefont {Chen}}]{Chen2017}%
  \BibitemOpen
  \bibfield  {author} {\bibinfo {author} {\bibfnamefont {P.}~\bibnamefont
  {Zhang}}, \bibinfo {author} {\bibfnamefont {I.}~\bibnamefont {Protsenko}},
  \bibinfo {author} {\bibfnamefont {V.}~\bibnamefont {Sandoghdar}}, \ and\
  \bibinfo {author} {\bibfnamefont {X.-W.}\ \bibnamefont {Chen}},\ }\bibfield
  {title} {\enquote {\bibinfo {title} {A single-emitter gain medium for bright
  coherent radiation from a plasmonic nanoresonator},}\ }\href {\doibase
  10.1021/acsphotonics.7b00608} {\bibfield  {journal} {\bibinfo  {journal} {ACS
  Photonics}\ }\textbf {\bibinfo {volume} {4}},\ \bibinfo {pages} {2738--2744}
  (\bibinfo {year} {2017}{\natexlab{a}})}\BibitemShut {NoStop}%
\bibitem [{\citenamefont {Mart\'{\i}n-Cano}\ \emph {et~al.}(2014)\citenamefont
  {Mart\'{\i}n-Cano}, \citenamefont {Haakh}, \citenamefont {Murr},\ and\
  \citenamefont {Agio}}]{Agio2014}%
  \BibitemOpen
  \bibfield  {author} {\bibinfo {author} {\bibfnamefont {D.}~\bibnamefont
  {Mart\'{\i}n-Cano}}, \bibinfo {author} {\bibfnamefont {H.~R.}\ \bibnamefont
  {Haakh}}, \bibinfo {author} {\bibfnamefont {K.}~\bibnamefont {Murr}}, \ and\
  \bibinfo {author} {\bibfnamefont {M.}~\bibnamefont {Agio}},\ }\bibfield
  {title} {\enquote {\bibinfo {title} {Large suppression of quantum
  fluctuations of light from a single emitter by an optical nanostructure},}\
  }\href {https://doi.org/10.1103/PhysRevLett.113.263605} {\bibfield  {journal}
  {\bibinfo  {journal} {Phys.\ Rev.\ Lett.}\ }\textbf {\bibinfo {volume}
  {113}},\ \bibinfo {pages} {263605} (\bibinfo {year} {2014})}\BibitemShut
  {NoStop}%
\bibitem [{\citenamefont {Haakh}\ and\ \citenamefont
  {Mart{\'{\i}}n-Cano}(2015)}]{Cano2015}%
  \BibitemOpen
  \bibfield  {author} {\bibinfo {author} {\bibfnamefont {H.~R.}\ \bibnamefont
  {Haakh}}\ and\ \bibinfo {author} {\bibfnamefont {D.}~\bibnamefont
  {Mart{\'{\i}}n-Cano}},\ }\bibfield  {title} {\enquote {\bibinfo {title}
  {Squeezed light from entangled nonidentical emitters via nanophotonic
  environments},}\ }\href {https://doi.org/10.1021/acsphotonics.5b00585}
  {\bibfield  {journal} {\bibinfo  {journal} {ACS Photonics}\ }\textbf
  {\bibinfo {volume} {2}},\ \bibinfo {pages} {1686} (\bibinfo {year}
  {2015})}\BibitemShut {NoStop}%
\bibitem [{\citenamefont {Ant\'{o}n}\ \emph {et~al.}(2017)\citenamefont
  {Ant\'{o}n}, \citenamefont {Maede-Razavi}, \citenamefont {Carre{\~{n}}o},
  \citenamefont {Thanopulos},\ and\ \citenamefont {Paspalakis}}]{Carreno2017}%
  \BibitemOpen
  \bibfield  {author} {\bibinfo {author} {\bibfnamefont {M.~A.}\ \bibnamefont
  {Ant\'{o}n}}, \bibinfo {author} {\bibfnamefont {S.}~\bibnamefont
  {Maede-Razavi}}, \bibinfo {author} {\bibfnamefont {F.}~\bibnamefont
  {Carre{\~{n}}o}}, \bibinfo {author} {\bibfnamefont {I.}~\bibnamefont
  {Thanopulos}}, \ and\ \bibinfo {author} {\bibfnamefont {E.}~\bibnamefont
  {Paspalakis}},\ }\bibfield  {title} {\enquote {\bibinfo {title} {Optical and
  microwave control of resonance fluorescence and squeezing spectra in a polar
  molecule},}\ }\href {https://doi.org/10.1103/PhysRevA.96.063812} {\bibfield
  {journal} {\bibinfo  {journal} {Phys. Rev. A}\ }\textbf {\bibinfo {volume}
  {96}},\ \bibinfo {pages} {063812} (\bibinfo {year} {2017})}\BibitemShut
  {NoStop}%
\bibitem [{\citenamefont {Lee}\ \emph {et~al.}(2013)\citenamefont {Lee},
  \citenamefont {Tame}, \citenamefont {Noh}, \citenamefont {Lim}, \citenamefont
  {Maier}, \citenamefont {Lee},\ and\ \citenamefont
  {Angelakis}}]{Angelakis2013}%
  \BibitemOpen
  \bibfield  {author} {\bibinfo {author} {\bibfnamefont {C.}~\bibnamefont
  {Lee}}, \bibinfo {author} {\bibfnamefont {M.}~\bibnamefont {Tame}}, \bibinfo
  {author} {\bibfnamefont {C.}~\bibnamefont {Noh}}, \bibinfo {author}
  {\bibfnamefont {J.}~\bibnamefont {Lim}}, \bibinfo {author} {\bibfnamefont
  {S.~A.}\ \bibnamefont {Maier}}, \bibinfo {author} {\bibfnamefont
  {J.}~\bibnamefont {Lee}}, \ and\ \bibinfo {author} {\bibfnamefont {D.~G.}\
  \bibnamefont {Angelakis}},\ }\bibfield  {title} {\enquote {\bibinfo {title}
  {Robust-to-loss entanglement generation using a quantum plasmonic
  nanoparticle array},}\ }\href
  {https://dx.doi.org/10.1088/1367-2630/15/8/083017} {\bibfield  {journal}
  {\bibinfo  {journal} {New J. Phys.}\ }\textbf {\bibinfo {volume} {15}},\
  \bibinfo {pages} {083017} (\bibinfo {year} {2013})}\BibitemShut {NoStop}%
\bibitem [{\citenamefont {Hou}\ \emph {et~al.}(2014)\citenamefont {Hou},
  \citenamefont {S\l{}owik}, \citenamefont {Lederer},\ and\ \citenamefont
  {Rockstuhl}}]{Rockstuhl2014b}%
  \BibitemOpen
  \bibfield  {author} {\bibinfo {author} {\bibfnamefont {J.}~\bibnamefont
  {Hou}}, \bibinfo {author} {\bibfnamefont {K.}~\bibnamefont {S\l{}owik}},
  \bibinfo {author} {\bibfnamefont {F.}~\bibnamefont {Lederer}}, \ and\
  \bibinfo {author} {\bibfnamefont {C.}~\bibnamefont {Rockstuhl}},\ }\bibfield
  {title} {\enquote {\bibinfo {title} {Dissipation-driven entanglement between
  qubits mediated by plasmonic nanoantennas},}\ }\href {\doibase
  10.1103/PhysRevB.89.235413} {\bibfield  {journal} {\bibinfo  {journal} {Phys.
  Rev. B}\ }\textbf {\bibinfo {volume} {89}},\ \bibinfo {pages} {235413}
  (\bibinfo {year} {2014})}\BibitemShut {NoStop}%
\bibitem [{\citenamefont {Otten}\ \emph {et~al.}(2015)\citenamefont {Otten},
  \citenamefont {Shah}, \citenamefont {Scherer}, \citenamefont {Min},
  \citenamefont {Pelton},\ and\ \citenamefont {Gray}}]{Otten2015}%
  \BibitemOpen
  \bibfield  {author} {\bibinfo {author} {\bibfnamefont {M.}~\bibnamefont
  {Otten}}, \bibinfo {author} {\bibfnamefont {R.~A.}\ \bibnamefont {Shah}},
  \bibinfo {author} {\bibfnamefont {N.~F.}\ \bibnamefont {Scherer}}, \bibinfo
  {author} {\bibfnamefont {M.}~\bibnamefont {Min}}, \bibinfo {author}
  {\bibfnamefont {M.}~\bibnamefont {Pelton}}, \ and\ \bibinfo {author}
  {\bibfnamefont {S.~K.}\ \bibnamefont {Gray}},\ }\bibfield  {title} {\enquote
  {\bibinfo {title} {Entanglement of two, three, or four plasmonically coupled
  quantum dots},}\ }\href {\doibase 10.1103/PhysRevB.92.125432} {\bibfield
  {journal} {\bibinfo  {journal} {Phys. Rev. B}\ }\textbf {\bibinfo {volume}
  {92}},\ \bibinfo {pages} {125432} (\bibinfo {year} {2015})}\BibitemShut
  {NoStop}%
\bibitem [{\citenamefont {Otten}\ \emph {et~al.}(2016)\citenamefont {Otten},
  \citenamefont {Larson}, \citenamefont {Min}, \citenamefont {Wild},
  \citenamefont {Pelton},\ and\ \citenamefont {Gray}}]{Otten2016}%
  \BibitemOpen
  \bibfield  {author} {\bibinfo {author} {\bibfnamefont {M.}~\bibnamefont
  {Otten}}, \bibinfo {author} {\bibfnamefont {J.}~\bibnamefont {Larson}},
  \bibinfo {author} {\bibfnamefont {M.}~\bibnamefont {Min}}, \bibinfo {author}
  {\bibfnamefont {S.~M.}\ \bibnamefont {Wild}}, \bibinfo {author}
  {\bibfnamefont {M.}~\bibnamefont {Pelton}}, \ and\ \bibinfo {author}
  {\bibfnamefont {S.~K.}\ \bibnamefont {Gray}},\ }\bibfield  {title} {\enquote
  {\bibinfo {title} {Origins and optimization of entanglement in plasmonically
  coupled quantum dots},}\ }\href {\doibase 10.1103/PhysRevA.94.022312}
  {\bibfield  {journal} {\bibinfo  {journal} {Phys. Rev. A}\ }\textbf {\bibinfo
  {volume} {94}},\ \bibinfo {pages} {022312} (\bibinfo {year}
  {2016})}\BibitemShut {NoStop}%
\bibitem [{\citenamefont {Hakami}\ and\ \citenamefont
  {Zubairy}(2016)}]{Zubairy2016}%
  \BibitemOpen
  \bibfield  {author} {\bibinfo {author} {\bibfnamefont {J.}~\bibnamefont
  {Hakami}}\ and\ \bibinfo {author} {\bibfnamefont {M.~S.}\ \bibnamefont
  {Zubairy}},\ }\bibfield  {title} {\enquote {\bibinfo {title}
  {Nanoshell-mediated robust entanglement between coupled quantum dots},}\
  }\href {\doibase 10.1103/PhysRevA.93.022320} {\bibfield  {journal} {\bibinfo
  {journal} {Phys. Rev. A}\ }\textbf {\bibinfo {volume} {93}},\ \bibinfo
  {pages} {022320} (\bibinfo {year} {2016})}\BibitemShut {NoStop}%
\bibitem [{\citenamefont {Straubel}\ \emph {et~al.}(2017)\citenamefont
  {Straubel}, \citenamefont {Sarniak}, \citenamefont {Rockstuhl},\ and\
  \citenamefont {S\l{}owik}}]{Slowik2017}%
  \BibitemOpen
  \bibfield  {author} {\bibinfo {author} {\bibfnamefont {J.}~\bibnamefont
  {Straubel}}, \bibinfo {author} {\bibfnamefont {R.}~\bibnamefont {Sarniak}},
  \bibinfo {author} {\bibfnamefont {C.}~\bibnamefont {Rockstuhl}}, \ and\
  \bibinfo {author} {\bibfnamefont {K.}~\bibnamefont {S\l{}owik}},\ }\bibfield
  {title} {\enquote {\bibinfo {title} {Entangled light from bimodal optical
  nanoantennas},}\ }\href {\doibase 10.1103/PhysRevB.95.085421} {\bibfield
  {journal} {\bibinfo  {journal} {Phys. Rev. B}\ }\textbf {\bibinfo {volume}
  {95}},\ \bibinfo {pages} {085421} (\bibinfo {year} {2017})}\BibitemShut
  {NoStop}%
\bibitem [{\citenamefont {Zhang}\ \emph
  {et~al.}(2017{\natexlab{b}})\citenamefont {Zhang}, \citenamefont {Zhao},
  \citenamefont {Gu}, \citenamefont {Chen}, \citenamefont {Hu},\ and\
  \citenamefont {Gong}}]{Gu2017b}%
  \BibitemOpen
  \bibfield  {author} {\bibinfo {author} {\bibfnamefont {F.}~\bibnamefont
  {Zhang}}, \bibinfo {author} {\bibfnamefont {D.}~\bibnamefont {Zhao}},
  \bibinfo {author} {\bibfnamefont {Y.}~\bibnamefont {Gu}}, \bibinfo {author}
  {\bibfnamefont {H.}~\bibnamefont {Chen}}, \bibinfo {author} {\bibfnamefont
  {X.}~\bibnamefont {Hu}}, \ and\ \bibinfo {author} {\bibfnamefont
  {Q.}~\bibnamefont {Gong}},\ }\bibfield  {title} {\enquote {\bibinfo {title}
  {Detuning-determined qubit-qubit entanglement mediated by plasmons: An
  effective model for dissipative systems},}\ }\href
  {https://dx.doi.org/10.1063/1.4984206} {\bibfield  {journal} {\bibinfo
  {journal} {J. Appl. Phys.}\ }\textbf {\bibinfo {volume} {121}},\ \bibinfo
  {pages} {203105} (\bibinfo {year} {2017}{\natexlab{b}})}\BibitemShut
  {NoStop}%
\bibitem [{\citenamefont {Iliopoulos}\ \emph {et~al.}(2017)\citenamefont
  {Iliopoulos}, \citenamefont {Terzis}, \citenamefont {Yannopapas},\ and\
  \citenamefont {Paspalakis}}]{Paspalakis2017}%
  \BibitemOpen
  \bibfield  {author} {\bibinfo {author} {\bibfnamefont {N.}~\bibnamefont
  {Iliopoulos}}, \bibinfo {author} {\bibfnamefont {A.~F.}\ \bibnamefont
  {Terzis}}, \bibinfo {author} {\bibfnamefont {V.}~\bibnamefont {Yannopapas}},
  \ and\ \bibinfo {author} {\bibfnamefont {E.}~\bibnamefont {Paspalakis}},\
  }\bibfield  {title} {\enquote {\bibinfo {title} {Prolonging entanglement
  dynamics near periodic plasmonic nanostructures},}\ }\href
  {https://doi.org/10.1103/PhysRevB.96.075405} {\bibfield  {journal} {\bibinfo
  {journal} {Phys. Rev. B}\ }\textbf {\bibinfo {volume} {96}},\ \bibinfo
  {pages} {075405} (\bibinfo {year} {2017})}\BibitemShut {NoStop}%
\bibitem [{\citenamefont {Hartsfield}\ \emph {et~al.}(2015)\citenamefont
  {Hartsfield}, \citenamefont {Chang}, \citenamefont {Yang}, \citenamefont
  {Ma}, \citenamefont {Shi}, \citenamefont {Sun}, \citenamefont {Shvets},
  \citenamefont {Link},\ and\ \citenamefont {Li}}]{Li2015}%
  \BibitemOpen
  \bibfield  {author} {\bibinfo {author} {\bibfnamefont {T.}~\bibnamefont
  {Hartsfield}}, \bibinfo {author} {\bibfnamefont {W.-S.}\ \bibnamefont
  {Chang}}, \bibinfo {author} {\bibfnamefont {S.-C.}\ \bibnamefont {Yang}},
  \bibinfo {author} {\bibfnamefont {T.}~\bibnamefont {Ma}}, \bibinfo {author}
  {\bibfnamefont {J.}~\bibnamefont {Shi}}, \bibinfo {author} {\bibfnamefont
  {L.}~\bibnamefont {Sun}}, \bibinfo {author} {\bibfnamefont {G.}~\bibnamefont
  {Shvets}}, \bibinfo {author} {\bibfnamefont {S.}~\bibnamefont {Link}}, \ and\
  \bibinfo {author} {\bibfnamefont {X.}~\bibnamefont {Li}},\ }\bibfield
  {title} {\enquote {\bibinfo {title} {Single quantum dot controls a plasmonic
  cavity¡¯s scattering and anisotropy},}\ }\href
  {http://dx.doi.org/10.1073/pnas.1508642112} {\bibfield  {journal} {\bibinfo
  {journal} {PNAS}\ }\textbf {\bibinfo {volume} {112}},\ \bibinfo {pages}
  {12288} (\bibinfo {year} {2015})}\BibitemShut {NoStop}%
\bibitem [{\citenamefont {Ding}\ \emph {et~al.}(2017)\citenamefont {Ding},
  \citenamefont {Li}, \citenamefont {Nan}, \citenamefont {Zhong}, \citenamefont
  {Zhou}, \citenamefont {Xiao}, \citenamefont {Wang},\ and\ \citenamefont
  {Zhang}}]{Zhang2017}%
  \BibitemOpen
  \bibfield  {author} {\bibinfo {author} {\bibfnamefont {S.-J.}\ \bibnamefont
  {Ding}}, \bibinfo {author} {\bibfnamefont {X.}~\bibnamefont {Li}}, \bibinfo
  {author} {\bibfnamefont {F.}~\bibnamefont {Nan}}, \bibinfo {author}
  {\bibfnamefont {Y.-T.}\ \bibnamefont {Zhong}}, \bibinfo {author}
  {\bibfnamefont {L.}~\bibnamefont {Zhou}}, \bibinfo {author} {\bibfnamefont
  {X.}~\bibnamefont {Xiao}}, \bibinfo {author} {\bibfnamefont {Q.-Q.}\
  \bibnamefont {Wang}}, \ and\ \bibinfo {author} {\bibfnamefont
  {Z.}~\bibnamefont {Zhang}},\ }\bibfield  {title} {\enquote {\bibinfo {title}
  {Strongly asymmetric spectroscopy in plasmon-exciton hybrid systems due to
  interference-induced energy repartitioning},}\ }\href {\doibase
  10.1103/PhysRevLett.119.177401} {\bibfield  {journal} {\bibinfo  {journal}
  {Phys. Rev. Lett.}\ }\textbf {\bibinfo {volume} {119}},\ \bibinfo {pages}
  {177401} (\bibinfo {year} {2017})}\BibitemShut {NoStop}%
\bibitem [{\citenamefont {Zhang}\ \emph
  {et~al.}(2017{\natexlab{c}})\citenamefont {Zhang}, \citenamefont {Meng},
  \citenamefont {Zhang}, \citenamefont {Luo}, \citenamefont {Yu}, \citenamefont
  {Yang}, \citenamefont {Zhang}, \citenamefont {Esteban}, \citenamefont
  {Aizpurua}, \citenamefont {Luo}, \citenamefont {Yang}, \citenamefont {Dong},\
  and\ \citenamefont {Hou}}]{Hou2017}%
  \BibitemOpen
  \bibfield  {author} {\bibinfo {author} {\bibfnamefont {Y.}~\bibnamefont
  {Zhang}}, \bibinfo {author} {\bibfnamefont {Q.-S.}\ \bibnamefont {Meng}},
  \bibinfo {author} {\bibfnamefont {L.}~\bibnamefont {Zhang}}, \bibinfo
  {author} {\bibfnamefont {Y.}~\bibnamefont {Luo}}, \bibinfo {author}
  {\bibfnamefont {Y.-J.}\ \bibnamefont {Yu}}, \bibinfo {author} {\bibfnamefont
  {B.}~\bibnamefont {Yang}}, \bibinfo {author} {\bibfnamefont {Y.}~\bibnamefont
  {Zhang}}, \bibinfo {author} {\bibfnamefont {R.}~\bibnamefont {Esteban}},
  \bibinfo {author} {\bibfnamefont {J.}~\bibnamefont {Aizpurua}}, \bibinfo
  {author} {\bibfnamefont {Y.}~\bibnamefont {Luo}}, \bibinfo {author}
  {\bibfnamefont {J.-L.}\ \bibnamefont {Yang}}, \bibinfo {author}
  {\bibfnamefont {Z.-C.}\ \bibnamefont {Dong}}, \ and\ \bibinfo {author}
  {\bibfnamefont {J.~G.}\ \bibnamefont {Hou}},\ }\bibfield  {title} {\enquote
  {\bibinfo {title} {Sub-nanometre control of the coherent interaction between
  a single molecule and a plasmonic nanocavity},}\ }\href
  {http://dx.doi.org/10.1038/ncomms15225} {\bibfield  {journal} {\bibinfo
  {journal} {Nat. Commun.}\ }\textbf {\bibinfo {volume} {8}},\ \bibinfo {pages}
  {15225} (\bibinfo {year} {2017}{\natexlab{c}})}\BibitemShut {NoStop}%
\bibitem [{\citenamefont {Chikkaraddy}\ \emph {et~al.}(2016)\citenamefont
  {Chikkaraddy}, \citenamefont {de~Nijs}, \citenamefont {Benz}, \citenamefont
  {Barrow}, \citenamefont {Scherman}, \citenamefont {Rosta}, \citenamefont
  {Demetriadou}, \citenamefont {Fox}, \citenamefont {Hess},\ and\ \citenamefont
  {Baumberg}}]{Baumberg2016}%
  \BibitemOpen
  \bibfield  {author} {\bibinfo {author} {\bibfnamefont {R.}~\bibnamefont
  {Chikkaraddy}}, \bibinfo {author} {\bibfnamefont {B.}~\bibnamefont
  {de~Nijs}}, \bibinfo {author} {\bibfnamefont {F.}~\bibnamefont {Benz}},
  \bibinfo {author} {\bibfnamefont {S.~J.}\ \bibnamefont {Barrow}}, \bibinfo
  {author} {\bibfnamefont {O.~A.}\ \bibnamefont {Scherman}}, \bibinfo {author}
  {\bibfnamefont {E.}~\bibnamefont {Rosta}}, \bibinfo {author} {\bibfnamefont
  {A.}~\bibnamefont {Demetriadou}}, \bibinfo {author} {\bibfnamefont
  {P.}~\bibnamefont {Fox}}, \bibinfo {author} {\bibfnamefont {O.}~\bibnamefont
  {Hess}}, \ and\ \bibinfo {author} {\bibfnamefont {J.~J.}\ \bibnamefont
  {Baumberg}},\ }\bibfield  {title} {\enquote {\bibinfo {title}
  {Single-molecule strong coupling at room temperature in plasmonic
  nanocavities},}\ }\href {http://dx.doi.org/10.1038/nature17974} {\bibfield
  {journal} {\bibinfo  {journal} {Nature (London)}\ }\textbf {\bibinfo {volume}
  {535}},\ \bibinfo {pages} {127} (\bibinfo {year} {2016})}\BibitemShut
  {NoStop}%
\bibitem [{\citenamefont {Liu}\ \emph {et~al.}(2017)\citenamefont {Liu},
  \citenamefont {Zhou}, \citenamefont {Yu}, \citenamefont {Zhang},
  \citenamefont {Wang}, \citenamefont {Liu}, \citenamefont {Wei}, \citenamefont
  {Chen},\ and\ \citenamefont {Wang}}]{Wang2017}%
  \BibitemOpen
  \bibfield  {author} {\bibinfo {author} {\bibfnamefont {R.}~\bibnamefont
  {Liu}}, \bibinfo {author} {\bibfnamefont {Z.-K.}\ \bibnamefont {Zhou}},
  \bibinfo {author} {\bibfnamefont {Y.-C.}\ \bibnamefont {Yu}}, \bibinfo
  {author} {\bibfnamefont {T.}~\bibnamefont {Zhang}}, \bibinfo {author}
  {\bibfnamefont {H.}~\bibnamefont {Wang}}, \bibinfo {author} {\bibfnamefont
  {G.}~\bibnamefont {Liu}}, \bibinfo {author} {\bibfnamefont {Y.}~\bibnamefont
  {Wei}}, \bibinfo {author} {\bibfnamefont {H.}~\bibnamefont {Chen}}, \ and\
  \bibinfo {author} {\bibfnamefont {X.-H.}\ \bibnamefont {Wang}},\ }\bibfield
  {title} {\enquote {\bibinfo {title} {Strong light-matter interactions in
  single open plasmonic nanocavities at the quantum optics limit},}\ }\href
  {\doibase 10.1103/PhysRevLett.118.237401} {\bibfield  {journal} {\bibinfo
  {journal} {Phys. Rev. Lett.}\ }\textbf {\bibinfo {volume} {118}},\ \bibinfo
  {pages} {237401} (\bibinfo {year} {2017})}\BibitemShut {NoStop}%
\bibitem [{\citenamefont {Scully}\ and\ \citenamefont
  {Zubairy}(1997)}]{Scully1997}%
  \BibitemOpen
  \bibfield  {author} {\bibinfo {author} {\bibfnamefont {M.~O.}\ \bibnamefont
  {Scully}}\ and\ \bibinfo {author} {\bibfnamefont {M.~S.}\ \bibnamefont
  {Zubairy}},\ }\href@noop {} {\emph {\bibinfo {title} {Quantum Optics}}}\
  (\bibinfo  {publisher} {Cambridge University Press},\ \bibinfo {address}
  {Cambridge},\ \bibinfo {year} {1997})\BibitemShut {NoStop}%
\bibitem [{\citenamefont {Reiserer}\ and\ \citenamefont
  {Rempe}(2015)}]{Rempe2015}%
  \BibitemOpen
  \bibfield  {author} {\bibinfo {author} {\bibfnamefont {A.}~\bibnamefont
  {Reiserer}}\ and\ \bibinfo {author} {\bibfnamefont {G.}~\bibnamefont
  {Rempe}},\ }\bibfield  {title} {\enquote {\bibinfo {title} {Cavity-based
  quantum networks with single atoms and optical photons},}\ }\href {\doibase
  10.1103/RevModPhys.87.1379} {\bibfield  {journal} {\bibinfo  {journal} {Rev.
  Mod. Phys.}\ }\textbf {\bibinfo {volume} {87}},\ \bibinfo {pages} {1379}
  (\bibinfo {year} {2015})}\BibitemShut {NoStop}%
\bibitem [{\citenamefont {Kira}\ \emph {et~al.}(2011)\citenamefont {Kira},
  \citenamefont {Koch}, \citenamefont {Smith}, \citenamefont {Hunter},\ and\
  \citenamefont {Cundiff}}]{Kira2011}%
  \BibitemOpen
  \bibfield  {author} {\bibinfo {author} {\bibfnamefont {M.}~\bibnamefont
  {Kira}}, \bibinfo {author} {\bibfnamefont {S.~W.}\ \bibnamefont {Koch}},
  \bibinfo {author} {\bibfnamefont {R.~P.}\ \bibnamefont {Smith}}, \bibinfo
  {author} {\bibfnamefont {A.~E.}\ \bibnamefont {Hunter}}, \ and\ \bibinfo
  {author} {\bibfnamefont {S.~T.}\ \bibnamefont {Cundiff}},\ }\bibfield
  {title} {\enquote {\bibinfo {title} {Quantum spectroscopy with
  schr{\"{o}}dinger-cat states},}\ }\href
  {https://dx.doi.org/10.1038/NPHYS2091} {\bibfield  {journal} {\bibinfo
  {journal} {Nat.\ Phys.}\ }\textbf {\bibinfo {volume} {7}},\ \bibinfo {pages}
  {799} (\bibinfo {year} {2011})}\BibitemShut {NoStop}%
\bibitem [{\citenamefont {Imamoglu}\ \emph {et~al.}(1997)\citenamefont
  {Imamoglu}, \citenamefont {Schmidt}, \citenamefont {Woods},\ and\
  \citenamefont {Deutsch}}]{Deutsch1997}%
  \BibitemOpen
  \bibfield  {author} {\bibinfo {author} {\bibfnamefont {A.}~\bibnamefont
  {Imamoglu}}, \bibinfo {author} {\bibfnamefont {H.}~\bibnamefont {Schmidt}},
  \bibinfo {author} {\bibfnamefont {G.}~\bibnamefont {Woods}}, \ and\ \bibinfo
  {author} {\bibfnamefont {M.}~\bibnamefont {Deutsch}},\ }\bibfield  {title}
  {\enquote {\bibinfo {title} {Strongly interacting photons in a nonlinear
  cavity},}\ }\href {\doibase 10.1103/PhysRevLett.79.1467} {\bibfield
  {journal} {\bibinfo  {journal} {Phys. Rev. Lett.}\ }\textbf {\bibinfo
  {volume} {79}},\ \bibinfo {pages} {1467} (\bibinfo {year}
  {1997})}\BibitemShut {NoStop}%
\bibitem [{\citenamefont {Birnbaum}\ \emph {et~al.}(2005)\citenamefont
  {Birnbaum}, \citenamefont {Boca}, \citenamefont {Miller}, \citenamefont
  {Boozer}, \citenamefont {Northup},\ and\ \citenamefont
  {Kimble}}]{Kimble2005}%
  \BibitemOpen
  \bibfield  {author} {\bibinfo {author} {\bibfnamefont {K.~M.}\ \bibnamefont
  {Birnbaum}}, \bibinfo {author} {\bibfnamefont {A.}~\bibnamefont {Boca}},
  \bibinfo {author} {\bibfnamefont {R.}~\bibnamefont {Miller}}, \bibinfo
  {author} {\bibfnamefont {A.~D.}\ \bibnamefont {Boozer}}, \bibinfo {author}
  {\bibfnamefont {T.~E.}\ \bibnamefont {Northup}}, \ and\ \bibinfo {author}
  {\bibfnamefont {H.~J.}\ \bibnamefont {Kimble}},\ }\bibfield  {title}
  {\enquote {\bibinfo {title} {Photon blockade in an optical cavity with one
  trapped atom},}\ }\href {http://dx.doi.org/10.1038/nature03804} {\bibfield
  {journal} {\bibinfo  {journal} {Nature (London)}\ }\textbf {\bibinfo {volume}
  {436}},\ \bibinfo {pages} {87} (\bibinfo {year} {2005})}\BibitemShut
  {NoStop}%
\bibitem [{\citenamefont {Rebi\ifmmode~\acute{c}\else \'{c}\fi{}}\ \emph
  {et~al.}(2002)\citenamefont {Rebi\ifmmode~\acute{c}\else \'{c}\fi{}},
  \citenamefont {Parkins},\ and\ \citenamefont {Tan}}]{Tan2002}%
  \BibitemOpen
  \bibfield  {author} {\bibinfo {author} {\bibfnamefont {S.}~\bibnamefont
  {Rebi\ifmmode~\acute{c}\else \'{c}\fi{}}}, \bibinfo {author} {\bibfnamefont
  {A.~S.}\ \bibnamefont {Parkins}}, \ and\ \bibinfo {author} {\bibfnamefont
  {S.~M.}\ \bibnamefont {Tan}},\ }\bibfield  {title} {\enquote {\bibinfo
  {title} {Photon statistics of a single-atom intracavity system involving
  electromagnetically induced transparency},}\ }\href {\doibase
  10.1103/PhysRevA.65.063804} {\bibfield  {journal} {\bibinfo  {journal} {Phys.
  Rev. A}\ }\textbf {\bibinfo {volume} {65}},\ \bibinfo {pages} {063804}
  (\bibinfo {year} {2002})}\BibitemShut {NoStop}%
\bibitem [{\citenamefont {Souza}\ \emph {et~al.}(2013)\citenamefont {Souza},
  \citenamefont {Figueroa}, \citenamefont {Chibani}, \citenamefont
  {Villas-Boas},\ and\ \citenamefont {Rempe}}]{Rempe2013}%
  \BibitemOpen
  \bibfield  {author} {\bibinfo {author} {\bibfnamefont {J.~A.}\ \bibnamefont
  {Souza}}, \bibinfo {author} {\bibfnamefont {E.}~\bibnamefont {Figueroa}},
  \bibinfo {author} {\bibfnamefont {H.}~\bibnamefont {Chibani}}, \bibinfo
  {author} {\bibfnamefont {C.~J.}\ \bibnamefont {Villas-Boas}}, \ and\ \bibinfo
  {author} {\bibfnamefont {G.}~\bibnamefont {Rempe}},\ }\bibfield  {title}
  {\enquote {\bibinfo {title} {Coherent control of quantum fluctuations using
  cavity electromagnetically induced transparency},}\ }\href
  {https://doi.org/10.1103/PhysRevLett.111.113602} {\bibfield  {journal}
  {\bibinfo  {journal} {Phys. Rev. Lett.}\ }\textbf {\bibinfo {volume} {111}},\
  \bibinfo {pages} {113602} (\bibinfo {year} {2013})}\BibitemShut {NoStop}%
\bibitem [{\citenamefont {Liew}\ and\ \citenamefont {Savona}(2010)}]{Liew2010}%
  \BibitemOpen
  \bibfield  {author} {\bibinfo {author} {\bibfnamefont {T.~C.~H.}\
  \bibnamefont {Liew}}\ and\ \bibinfo {author} {\bibfnamefont {V.}~\bibnamefont
  {Savona}},\ }\bibfield  {title} {\enquote {\bibinfo {title} {Single photons
  from coupled quantum modes},}\ }\href {\doibase
  10.1103/PhysRevLett.104.183601} {\bibfield  {journal} {\bibinfo  {journal}
  {Phys. Rev. Lett.}\ }\textbf {\bibinfo {volume} {104}},\ \bibinfo {pages}
  {183601} (\bibinfo {year} {2010})}\BibitemShut {NoStop}%
\bibitem [{\citenamefont {Bamba}\ \emph {et~al.}(2011)\citenamefont {Bamba},
  \citenamefont {Imamoglu}, \citenamefont {Carusotto},\ and\ \citenamefont
  {Ciuti}}]{Ciuti2011}%
  \BibitemOpen
  \bibfield  {author} {\bibinfo {author} {\bibfnamefont {M.}~\bibnamefont
  {Bamba}}, \bibinfo {author} {\bibfnamefont {A.}~\bibnamefont {Imamoglu}},
  \bibinfo {author} {\bibfnamefont {I.}~\bibnamefont {Carusotto}}, \ and\
  \bibinfo {author} {\bibfnamefont {C.}~\bibnamefont {Ciuti}},\ }\bibfield
  {title} {\enquote {\bibinfo {title} {Origin of strong photon antibunching in
  weakly nonlinear photonic molecules},}\ }\href {\doibase
  10.1103/PhysRevA.83.021802} {\bibfield  {journal} {\bibinfo  {journal} {Phys.
  Rev. A}\ }\textbf {\bibinfo {volume} {83}},\ \bibinfo {pages} {021802}
  (\bibinfo {year} {2011})}\BibitemShut {NoStop}%
\bibitem [{\citenamefont {Bohren}\ and\ \citenamefont
  {Huffman}(1983)}]{Bohren1983}%
  \BibitemOpen
  \bibfield  {author} {\bibinfo {author} {\bibfnamefont {C.~F.}\ \bibnamefont
  {Bohren}}\ and\ \bibinfo {author} {\bibfnamefont {D.~R.}\ \bibnamefont
  {Huffman}},\ }\href@noop {} {\emph {\bibinfo {title} {Absorption and
  Scattering of Light by Small Particles}}}\ (\bibinfo  {publisher} {Wiley},\
  \bibinfo {address} {New York},\ \bibinfo {year} {1983})\BibitemShut {NoStop}%
\bibitem [{\citenamefont {Johansson}\ \emph {et~al.}(2012)\citenamefont
  {Johansson}, \citenamefont {Nation},\ and\ \citenamefont {Nori}}]{Qutip2012}%
  \BibitemOpen
  \bibfield  {author} {\bibinfo {author} {\bibfnamefont {J.~R.}\ \bibnamefont
  {Johansson}}, \bibinfo {author} {\bibfnamefont {P.~D.}\ \bibnamefont
  {Nation}}, \ and\ \bibinfo {author} {\bibfnamefont {F.}~\bibnamefont
  {Nori}},\ }\bibfield  {title} {\enquote {\bibinfo {title} {Qutip: An
  open-source python framework for the dynamics of open quantum systems},}\
  }\href {http://dx.doi.org/10.1016/j.cpc.2012.02.021} {\bibfield  {journal}
  {\bibinfo  {journal} {Comp.\ Phys.\ Comm.}\ }\textbf {\bibinfo {volume}
  {183}},\ \bibinfo {pages} {1760} (\bibinfo {year} {2012})}\BibitemShut
  {NoStop}%
\bibitem [{\citenamefont {Johansson}\ \emph {et~al.}(2013)\citenamefont
  {Johansson}, \citenamefont {Nation},\ and\ \citenamefont {Nori}}]{Qutip2013}%
  \BibitemOpen
  \bibfield  {author} {\bibinfo {author} {\bibfnamefont {J.~R.}\ \bibnamefont
  {Johansson}}, \bibinfo {author} {\bibfnamefont {P.~D.}\ \bibnamefont
  {Nation}}, \ and\ \bibinfo {author} {\bibfnamefont {F.}~\bibnamefont
  {Nori}},\ }\bibfield  {title} {\enquote {\bibinfo {title} {Qutip 2: A python
  framework for the dynamics of open quantum systems},}\ }\href
  {https://doi.org/10.1016/j.cpc.2012.11.019} {\bibfield  {journal} {\bibinfo
  {journal} {Comp.\ Phys.\ Comm.}\ }\textbf {\bibinfo {volume} {184}},\
  \bibinfo {pages} {1234} (\bibinfo {year} {2013})}\BibitemShut {NoStop}%
\bibitem [{\citenamefont {Waks}\ and\ \citenamefont
  {Sridharan}(2010)}]{Waks2010}%
  \BibitemOpen
  \bibfield  {author} {\bibinfo {author} {\bibfnamefont {E.}~\bibnamefont
  {Waks}}\ and\ \bibinfo {author} {\bibfnamefont {D.}~\bibnamefont
  {Sridharan}},\ }\bibfield  {title} {\enquote {\bibinfo {title} {Cavity qed
  treatment of interactions between a metal nanoparticle and a dipole
  emitter},}\ }\href {\doibase 10.1103/PhysRevA.82.043845} {\bibfield
  {journal} {\bibinfo  {journal} {Phys. Rev. A}\ }\textbf {\bibinfo {volume}
  {82}},\ \bibinfo {pages} {043845} (\bibinfo {year} {2010})}\BibitemShut
  {NoStop}%
\bibitem [{\citenamefont {Zhang}\ \emph {et~al.}(2006)\citenamefont {Zhang},
  \citenamefont {Govorov},\ and\ \citenamefont {Bryant}}]{Bryant2006}%
  \BibitemOpen
  \bibfield  {author} {\bibinfo {author} {\bibfnamefont {W.}~\bibnamefont
  {Zhang}}, \bibinfo {author} {\bibfnamefont {A.~O.}\ \bibnamefont {Govorov}},
  \ and\ \bibinfo {author} {\bibfnamefont {G.~W.}\ \bibnamefont {Bryant}},\
  }\bibfield  {title} {\enquote {\bibinfo {title} {Semiconductor-metal
  nanoparticle molecules: Hybrid excitons and the nonlinear fano effect},}\
  }\href {\doibase 10.1103/PhysRevLett.97.146804} {\bibfield  {journal}
  {\bibinfo  {journal} {Phys. Rev. Lett.}\ }\textbf {\bibinfo {volume} {97}},\
  \bibinfo {pages} {146804} (\bibinfo {year} {2006})}\BibitemShut {NoStop}%
\bibitem [{\citenamefont {Cohen-Tannoudji}\ and\ \citenamefont
  {Reynaud}(1977)}]{Cohen1977}%
  \BibitemOpen
  \bibfield  {author} {\bibinfo {author} {\bibfnamefont {C.}~\bibnamefont
  {Cohen-Tannoudji}}\ and\ \bibinfo {author} {\bibfnamefont {S.}~\bibnamefont
  {Reynaud}},\ }\bibfield  {title} {\enquote {\bibinfo {title} {Dressed-atom
  description of resonance fluorescence and absorption spectra of a multi-level
  atom in an intense laser beam},}\ }\href
  {http://stacks.iop.org/0022-3700/10/i=3/a=005} {\bibfield  {journal}
  {\bibinfo  {journal} {J. Phys. B: Atom. Molec. Phys.}\ }\textbf {\bibinfo
  {volume} {10}},\ \bibinfo {pages} {345} (\bibinfo {year} {1977})}\BibitemShut
  {NoStop}%
\bibitem [{\citenamefont {Carmichael}\ \emph {et~al.}(1991)\citenamefont
  {Carmichael}, \citenamefont {Brecha},\ and\ \citenamefont
  {Rice}}]{Carmichael1991}%
  \BibitemOpen
  \bibfield  {author} {\bibinfo {author} {\bibfnamefont {H.~J.}\ \bibnamefont
  {Carmichael}}, \bibinfo {author} {\bibfnamefont {R.~J.}\ \bibnamefont
  {Brecha}}, \ and\ \bibinfo {author} {\bibfnamefont {P.~R.}\ \bibnamefont
  {Rice}},\ }\bibfield  {title} {\enquote {\bibinfo {title} {Quantum
  interference and collapse of the wavefunction in cavity qed},}\ }\href
  {\doibase https://doi.org/10.1016/0030-4018(91)90194-I} {\bibfield  {journal}
  {\bibinfo  {journal} {Opt. Commun.}\ }\textbf {\bibinfo {volume} {82}},\
  \bibinfo {pages} {73} (\bibinfo {year} {1991})}\BibitemShut {NoStop}%
\end{thebibliography}%

\end{document}